s

# Neyman-Pearson (NP) Classification Algorithms and NP Receiver Operating Characteristics (NP-ROC)

**(Short title: NP Classification and NP-ROC)**

Xin Tong,[1†*] Yang Feng,[2†] Jingyi Jessica Li[3*]

[1] Department of Data Sciences and Operations, Marshall School of Business, University of Southern California, Los Angeles, CA, United States.
[2] Department of Statistics, Columbia University, New York, NY, United States.
[3] Department of Statistics, University of California, Los Angeles, CA, United States.
[†] These authors contributed equally to this work.
[*] To whom correspondence should be addressed: E-mails: xint@marshall.usc.edu and jli@stat.ucla.edu

## Abstract

In many binary classification applications, such as disease diagnosis and spam detection, practitioners commonly face the need to limit type I error (i.e., the conditional probability of misclassifying a class 0 observation as class 1) so that it remains below a desired threshold. To address this need, the Neyman-Pearson (NP) classification paradigm is a natural choice; it minimizes type II error (i.e., the conditional probability of misclassifying a class 1 observation as class 0) while enforcing an upper bound, $\alpha$, on the type I error. Although the NP paradigm has a century-long history in hypothesis testing, it has not been well recognized and implemented in classification schemes. Common practices that directly limit the empirical type I error to no more than $\alpha$ do not satisfy the type I error control objective because the resulting classifiers are still likely to have type I errors much larger than $\alpha$. As a result, the NP paradigm has not been properly implemented for many classification scenarios in practice. In this work, we develop the first umbrella algorithm that implements the NP paradigm for all scoring-type classification methods, including popular methods such as logistic regression, support vector machines and random forests. Powered by this umbrella algorithm, we propose a novel graphical tool for NP classification methods: NP receiver operating characteristic (NP-ROC) bands, motivated by the popular receiver operating characteristic (ROC) curves. NP-ROC bands will help choose $\alpha$ in a data adaptive way and compare different NP classifiers. We demonstrate the use and properties of the NP umbrella algorithm and NP-ROC bands, available in the `R` package `nproc`, through simulation and real data case studies.

## Introduction

In statistics and machine learning, the purpose of classification is to automatically predict discrete outcomes (i.e., class labels) for new observations on the basis of labeled training data. The development of classification theory, methods and applications has been a dynamic area of research for more than half a century (*1*). Well known examples include disease diagnosis, email spam filters, and image classification. Binary classification, in which the class labels are 0 and 1, is the most common type. Most binary classifiers are constructed to minimize the expected classification error (i.e., *risk*), which is a weighted sum of type I and type II errors. We refer to this paradigm as the classical classification paradigm in this paper. *Type I error* is defined as the conditional probability of misclassifying a class 0 observation as a class 1 observation; it is also called the *false positive rate* (i.e., 1 − *specificity*). *Type II error* is the conditional probability of misclassifying a class 1 observation as class 0; it is also called the *false negative rate* (i.e., 1 − *sensitivity*). Note that in some specific scientific or medical contexts, positiveness and negativeness have strict definitions, and their definition of "false positive rate" may be different from the probability of classifying an observation whose true label is 0 as class 1. As the binary class labels can be arbitrarily defined and switched, in the following text we refer to the prioritized type of error as the type I error.

In the risk, the weights of the type I and type II errors are the marginal probabilities of classes 0 and 1, respectively. In real-world applications, however, users' priorities for type I and type II errors may differ from these weights. For example, in cancer diagnosis, a type I error (i.e., misdiagnosing a cancer patient as healthy) has more severe consequences than a type II error (i.e., misdiagnosing a healthy patient with cancer); the latter may lead to extra medical costs and patient anxiety but will not result in tragic loss of life (*2–5*). For such applications, a prioritized control of asymmetric classification errors is sorely needed. The Neyman-Pearson (NP) classification paradigm was developed for this purpose (*6–9*); it seeks a classifier that minimizes the type II error while maintaining the type I error below a user-specified level $\alpha$, usually a small value (e.g., 5%). In statistical learning theory, this target classifier is called the oracle NP classifier; it achieves the minimum type II error given an upper bound $\alpha$ on the type I error. It is different from the cost-sensitive learning (*10*, *11*) and the asymmetric support vector machines (*12*), which also address asymmetric classification errors but provide no probabilistic control on the errors. Previous studies addressed NP classification using both empirical risk minimization (ERM) (*6–8*, *13–15*) and plug-in approaches (*9*, *16*). For a review of the current status of NP classification, we refer the readers to (*17*). A main advantage of the NP classification is that it is a general framework that allows users to find classifiers with (population) type I errors under $\alpha$ with high probability. In many biomedical, engineering and social applications, users often have pre-specified $\alpha$ values to reflect their tolerance on the type I errors and use diverse classification algorithms. Example applications include diagnosis of coronary artery disease (*18*), cancer early warning system (*19*), network security control (*20*, *21*), Environmental Protection Agency water security research (*22*), prediction of regional and international conflicts (*23*, *24*) and the Early Warning Project to identify countries at risk of new mass atrocities (*25*). However, existing ad-hoc use of classification algorithms cannot control type I errors under $\alpha$ with high probability. While the NP paradigm can address the needs, how to implement it with diverse classification algorithms remains a challenge.

In this paper, we address two important but unanswered questions regarding the practicality of the NP classification paradigm and the evaluation of NP classification methods. The first question is how to adapt popular classification methods (e.g., logistic regression (*26*), support vector machines (*27*), AdaBoost (*28*), and random forests (*29*)) to construct NP classifiers. We address this question by proposing an umbrella algorithm to implement a broad class of classification methods under the NP paradigm. The second question is how to evaluate and compare the performance of different NP classification methods. We propose NP-ROC bands, a variant of ROC, as a new visualization tool for NP classification. Possible use of NP-ROC bands for practitioners include but not limited to (I) choosing $\alpha$ in a data adaptive way and (II) comparing different NP classifiers.

## Mathematical Formulation

To facilitate our discussion of technical details, we introduce the following mathematical notation and review our previous theoretic formulation (*8*, *9*, *16*) to further explain the classical and NP classification paradigms. Let $(X, Y)$ be random variables where $X \in \mathcal{X} \subset \mathbb{R}^d$ is a vector of $d$ features, and $Y \in \{0,1\}$ represents a binary class label. A data set that contains independent observations $\{(x_i, y_i)\}_{i=1}^n$ sampled from the joint distribution of $(X, Y)$ is often divided into training data and test data. Based on training data, a classifier $\phi(\cdot)$ is a mapping $\phi: \mathcal{X} \to \{0,1\}$ that returns the predicted class label given *X*. Classification errors occur when $\phi(X) \neq Y$, and the binary loss is defined as $I(\phi(X) \neq Y)$, where $I(\cdot)$ denotes the indicator function. The risk is defined as $R(\phi) = \mathbb{E}[I(\phi(X) \neq Y)] = \mathbb{P}(\phi(X) \neq Y)$, which can be expressed as a weighted sum of type I and II errors: $R(\phi) = \mathbb{P}(Y = 0)R_0(\phi) + \mathbb{P}(Y = 1)R_1(\phi)$, where $R_0(\phi) = \mathbb{P}(\phi(X) \neq Y | Y = 0)$ denotes the (population) type I error, and $R_1(\phi) = \mathbb{P}(\phi(X) \neq Y | Y = 1)$ denotes the (population) type II error. The classical classification paradigm aims to mimic the *classical oracle classifier* $\phi^*$ that minimizes the risk,

$$\phi^* = \underset{\phi}{\operatorname{argmin}}\, R(\phi).$$

In contrast, the NP classification paradigm aims to mimic the *NP oracle classifier* $\phi_\alpha^*$ with respect to a pre-specified upper bound on the type I error, $\alpha$,

$$\phi_\alpha^* = \underset{\phi: R_0(\alpha) \le \alpha}{\operatorname{argmin}}\; R_1(\phi),$$

where $\alpha$ reflects users' conservative attitude (i.e., priority) towards type I error. Fig. 1 shows a toy example that demonstrates the difference between a classical oracle classifier that minimizes the risk and an NP oracle classifier that minimizes the type II error given type I error $\le \alpha = 0.05$.

In practice, $R(\cdot)$, $R_0(\cdot)$ and $R_1(\cdot)$ are unobservable because they depend on the unknown joint distribution of $(X, Y)$. Instead, their estimates based on data (i.e., the empirical risk, empirical type I error and empirical type II error) are often used in practice. Here we denote the empirical risk and type I and II errors based on training data as $\hat{R}(\cdot)$, $\hat{R}_0(\cdot)$ and $\hat{R}_1(\cdot)$ and the empirical risk and type I and II errors based on test data as $\tilde{R}(\cdot)$, $\tilde{R}_0(\cdot)$ and $\tilde{R}_1(\cdot)$.

Due to the wide applications of classification in real-world problems, a vast array of classification methods have been developed to construct "good" binary classifiers. In this paper, we focus on the scoring type of classification methods, which first train a scoring function $f: \mathcal{X} \to \mathbb{R}$ using the training data. The scoring function $f(\cdot)$ assigns a classification score $f(x)$ to an observation $x \in \mathbb{R}^d$. By setting a threshold $c \in \mathbb{R}$ on the classification scores, a classifier can be obtained. In other words, we consider classifiers with the form $\phi^c(\cdot) = I(f(\cdot) > c)$. Most popular classification methods are of this type (*30*). For example, logistic regression, support vector machines, naïve Bayes and neural networks all output a numeric value, i.e., a classification score, to represent the degree to which a test data point belongs to class 1. The classification scores can be strict probabilities or uncalibrated numeric values, as long as a higher score indicates a higher probability of an observation belonging to class 1. Another type of classification algorithms (e.g., random forests) use bagging to generate an ensemble of classifiers, each of which predicts a class label for a test data point; in these scenarios, the proportion of predicted labels being 1 serves as a classification score.

## An Umbrella Algorithm for NP Classification

Recent studies describe several NP classifiers that respect the type I error bound with high probability. They are built upon plug-in estimators of density ratios (i.e., class 1 density / class 0 density) (*9*, *16*). However, these classifiers are only applicable under a number of restricted scenarios, such as low feature dimension (*9*) and feature independence (*16*). Many other statistical and machine learning algorithms have been shown to be effective classification methods but not yet implemented under the NP paradigm; these include but not limited to (penalized) logistic regression, support vector machines, AdaBoost, and random forests. To develop NP classifiers that are diverse and adaptable to various practical scenarios and avoid duplicating efforts, we choose to implement these popular classification algorithms under the NP paradigm rather than construct numerous new NP classifiers based on complex models for density ratios. Moreover, classifiers in (*9*) use VC theory to guide an upper bound for empirical type I error and require a sample size larger than available in many modern applications, while classifiers in (*16*) resort to concentration inequalities to get an explicit order at the expense of possible overly conservatism in type I errors. Here we develop an alternative approach by calculating exact probabilities (under mild continuity assumptions) based on order statistic distributions to find thresholds on classification scores under the NP paradigm.

The first main contribution of this paper is our proposed umbrella algorithm that adapts popular classification methods to the NP paradigm. These methods include logistic regression (LR) and penalized LR (penLR), support vector machines (SVM), linear discriminant analysis (LDA), naïve Bayes (NB), AdaBoost, classification trees, and random forests (RF). Specifically, we seek an efficient way to choose a threshold for the classification scores predicted by each algorithm so that the threshold leads to classifiers with type I errors below the user-specified upper bound $\alpha$ with high probability. Such an algorithm is needed because the naïve approach, which simply picks a threshold by setting the empirical type I error to no more than $\alpha$, fails to satisfy the type I error constraint, as demonstrated in the simulation study described below.

**Simulation 1.** Data are generated from two Gaussian distributions: $(X|Y=0) \sim N(0,1)$ and $(X|Y=1) \sim N(2,1)$, with $\mathbb{P}(Y=0)=\mathbb{P}(Y=1)=0.5$. We denote a data set from this distribution as $\{(x_i, y_i)\}_{i=1}^N$, where $N=1000$. The classifiers we consider are $I(X>c)$, where $c \in \mathbb{R}$. Here the classification scoring function $f(x)=x$, the identity function, because in this one-dimensional case, $x_i$'s naturally serve as good classification scores. Hence, no training for $f$ is needed, and we only rely on this data set to find $c$ such that the corresponding classifier has the population type I error below $\alpha$ with high probability. Fig. 2 illustrates this problem. The black curves denote the oracle ROC curves, which trace the population type I error and (1 − population type II error) of these classifiers as $c$ varies. To find a value of $c$ such that the corresponding classifier has type I error $\leq \alpha = 0.05$, a common and intuitive practice is to choose the smallest $c$ such that the empirical type I error is no greater than $\alpha$, resulting in a classifier $\bar{\phi}_\alpha(\cdot) = I(\cdot > \bar{c}_\alpha)$, where $\bar{c}_\alpha = \inf\left\{c: \frac{\sum_{i=1}^N I(x_i>c, y_i=0)}{\sum_{i=1}^N I(y_i=0)} \leq \alpha\right\}$. In our simulation, we generate $D=1000$ data sets; this procedure results in $D$ classifiers, $\bar{\phi}_\alpha^{(1)}, \cdots, \bar{\phi}_\alpha^{(D)}$, shown as red "×"s on the oracle ROC curve (Fig. 2A). However, only approximately half of these classifiers have type I errors below $\alpha$, which is far from achieving the users' goal. Therefore, this commonly used method does not work well in this case. We also use another approach based on the 5-fold cross-validation (CV), though in this case no training is needed to estimate the type I error of the classifier $I(X>c)$ for every $c \in \mathbb{R}$. Concretely, we randomly split the class 0 data into five folds. On each fold we calculate the empirical type I error of the classifier, and we take the average of the five empirical type I errors as the CV type I error, denoted by $\hat{R}_0^{CV}(c)$. Based on these CV type I errors of various $c$, we can construct a classifier $\tilde{\phi}_\alpha(\cdot) = I(\cdot > \tilde{c}_\alpha)$, where $\tilde{c}_\alpha = \inf\{c: \hat{R}_0^{CV}(c) \le \alpha\}$. Applying this procedure to the $D = 1000$ data sets, we obtain $D$ classifiers, $\tilde{\phi}_\alpha^{(1)}, \cdots, \tilde{\phi}_\alpha^{(D)}$, shown as cyan "×"s on the oracle ROC curve (Fig. 2B). However, still only about half of these classifiers have type I errors below $\alpha$. Therefore, this cross-validation based approach still does not work well in this case.

In view of this failure, we propose an umbrella algorithm to implement the NP paradigm, summarized as pseudocodes below.

---

**Algorithm 1.** An NP umbrella algorithm

---

**input**:

training data: a mixed i.i.d. sample $\mathcal{S} = \mathcal{S}^0 \cup \mathcal{S}^1$, where $\mathcal{S}^0$ and $\mathcal{S}^1$ are class 0 and class 1 samples respectively

$\alpha$: type I error upper bound, $0 \le \alpha \le 1$; [default $\alpha = 0.05$]

$\delta$: a small tolerance level, $0 < \delta < 1$; [default $\delta = 0.05$]

$M$: number of random splits on $\mathcal{S}^0$; [default $M = 1$]

**function** RankThreshold$(n, \alpha, \delta)$

  **for** $k$ in $\{1, \cdots, n\}$ **do** ◃ *for each rank threshold candidate k*

    $v(k) \leftarrow \sum_{j=k}^{n} \binom{n}{j} (1-\alpha)^j \alpha^{n-j}$ ◃ *calculate the violation rate upper bound*

  $k^* \leftarrow \min\{k \in \{1, \cdots, n\}: v(k) \le \delta\}$ ◃ *pick the rank threshold*

  **return** $k^*$

**procedure** NPclassifier$(\mathcal{S}, \alpha, \delta, M)$

  $n = \lceil |\mathcal{S}^0|/2 \rceil$ ◃ *denote half of the size of* $|\mathcal{S}^0|$ *as* $n$

  $k^* \leftarrow$ RankThreshold$(n, \alpha, \delta)$ ◃ *find the rank threshold*

  **for** $i$ in $\{1, \cdots, M\}$ **do** ◃ *randomly split* $\mathcal{S}^0$ *for* $M$ *times*

    $\mathcal{S}_{i,1}^0, \mathcal{S}_{i,2}^0 \leftarrow$ random split on $\mathcal{S}^0$ ◃ *each time randomly split* $\mathcal{S}^0$ *into two halves with equal sizes*

    $\mathcal{S}_i \leftarrow \mathcal{S}_{i,1}^0 \cup \mathcal{S}^1$ ◃ *combine* $\mathcal{S}_{i,1}^0$ *and* $\mathcal{S}^1$

    $\mathcal{S}_{i,2}^0 = \{x_1, \cdots, x_n\}$ ◃ *write* $\mathcal{S}_{i,2}^0$ *as a set of* $n$ *data points*

    $f_i \leftarrow$ `ClassificationAlgorithm`$(\mathcal{S}_i)$ ◃ *train a scoring function* $f_i$ *on* $\mathcal{S}_i$

    $\mathcal{T}_i = \{t_{i,1}, \cdots, t_{i,n}\} \leftarrow \{f_i(x_1), \cdots, f_i(x_n)\}$ ◃ *apply the scoring function* $f_i$ *to* $\mathcal{S}_{i,2}^0$ *to obtain a set of score threshold candidates*

    $\{t_{i,(1)}, \cdots, t_{i,(n)}\} \leftarrow$ `sort`$(\mathcal{T}_i)$ ◃ *sort elements of* $\mathcal{T}_i$ *in an increasing order*

    $t_i^* \leftarrow t_{i,(k^*)}$ ◃ *find the score threshold corresponding to the rank threshold* $k^*$

    $\phi_i(X) = I(f_i(X) > t_i^*)$ ◃ *construct an NP classifier based on the scoring function* $f_i$ *and the threshold* $t_i^*$

**output:**

an ensemble NP classifier $\hat{\phi}_\alpha(X) = I\left(\frac{1}{M}\sum_{i=1}^{M} \phi_i(X) \ge \frac{1}{2}\right)$ ◃ *by majority vote*

---

The essential idea is to choose the smallest threshold on the classification scores such that the violation rate (i.e., the probability that the population type I error exceeds $\alpha$) is controlled under some pre-specified tolerance parameter $\delta$, i.e., $\mathbb{P}[R_0(\phi^c) > \alpha] \le \delta$. The threshold $c$ is to be chosen from an order statistic of classification scores of a left-out class 0 sample, which is not used to train base algorithms (e.g., logistic regression, support vector machines, and random forests) to obtain $f$. Because we do not impose any assumptions on the underlying data-generating process, it is not feasible to establish oracle-type theoretical properties regarding type II errors under this umbrella algorithm. However, because users may favor different base algorithms, this umbrella algorithm is generally applicable in light of the general preference towards conservatism with regard to type I errors. The theoretical foundation of the umbrella algorithm is explained by the following proposition.

**Proposition 1.** *Suppose that we divide the training data into two parts, one with data from both classes 0 and 1 for training a base algorithm (e.g., logistic regression) to obtain $f$ and the other as a left-out class 0 sample for choosing $c$. Applying the resulting $f$ to the left-out class 0 sample of size $n$, we denote the resulting classification scores as $T_1, \cdots, T_n$, which are real-valued random variables. Then, we denote by $T_{(k)}$ the $k$-th order statistic (i.e., $T_{(1)} \le \cdots \le T_{(n)}$). For a new observation, if we denote its classification score based on $f$ as $T$, we can construct a classifier $\hat{\phi}_k = (T > T_{(k)})$. Then, the population type I error of $\hat{\phi}_k$, denoted by $R_0(\hat{\phi}_k)$, is a function of $T_{(k)}$ and hence a random variable. Assuming the data used to train the base algorithm and the left-out class 0 data are independent, we have*

$$\mathbb{P}\left[R_0(\hat{\phi}_k) > \alpha\right] \le \sum_{j=k}^{n} \binom{n}{j} (1-\alpha)^j \alpha^{n-j} . \tag{1}$$

*That is, the probability that the type I error of $\hat{\phi}_k$ exceeds $\alpha$ is under a constant that only depends on $k$ and $\alpha$. We call this probability the "violation rate" of $\hat{\phi}_k$ and denote its upper bound by $v(k) = \sum_{j=k}^{n} \binom{n}{j} (1-\alpha)^j \alpha^{n-j}$. When $T_i$'s are continuous, this bound is tight.*

For the proof of Proposition 1, please refer to the Supplementary Materials. Note that Proposition 1 is general, as it does not rely on any distributional assumptions or on base algorithm characteristics.

Clearly $v(k)$ decreases as $k$ increases. If we would like to construct an NP classifier based on an order statistic of the classification scores of the left-out class 0 sample, the right order should be

$$k^* = \min\{k \in \{1, \cdots, n\}: v(k) \le \delta\}. \tag{2}$$

To control the violation rate under $\delta$ at least in the extreme case when $k = n$, we need to have $v(n) = (1-\alpha)^n \le \delta$. If the $n$-th order statistic cannot guarantee this violation rate control, other order statistics certainly cannot. Therefore, given $\alpha$ and $\delta$, we need to have the minimum sample size requirement $n \ge \log \delta / \log(1-\alpha)$ for type I error violation rate control; otherwise, the control cannot be achieved, at least by this order statistic approach.

Algorithm 1 describes our umbrella NP algorithm, which supports popular classification methods such as LR and SVM. Proposition 1 provides the theoretical guarantee on type I error violation rate control for one random split ($M = 1$ in the algorithm). With multiple ($M > 1$) random splits, with each split dividing class 0 training data into two halves, an ensemble classifier by majority voting is demonstrated to maintain the type I error control and reduce the expectation and variance of the type II error in numerical experiments (Simulation S2 in Supplementary Materials).

Applying this algorithm to the example in Simulation 1, with $\alpha = 0.05$, $\delta = 0.05$ and the number of random splits $M = 1$, we construct $D$ NP classifiers $\hat{\phi}_\alpha^{(1)}, \dots, \hat{\phi}_\alpha^{(D)}$ based on the $D = 1000$ data sets. We mark these $D$ NP classifiers on the oracle ROC curve (shown as blue "+"s in Fig. 2C). Unlike the classifiers constructed by the naïve approach (red "×"s in Fig. 2A) or the 5-fold cross-validation (cyan "×"s in Fig. 2B), we see that these $D$ NP classifiers have type I errors below $\alpha$ with high (at least $1 - \delta$) probability.

## Empirical ROC Curves

A popular tool for evaluating binary classification methods is the receiver operating characteristic (ROC) curve, which provides graphical illustration of the overall performance of a classification method, by showing its all possible type I and type II errors. ROC curves have numerous applications in signal detection theory, diagnostic systems, and medical decision making, among other fields (*31–33*). For the scoring type of binary classification methods we focus on in this paper, ROC curves illustrate the overall performance at all possible values of the threshold $c$ on the output classification scores. An ROC space is defined as a two dimensional $[0,1] \times [0,1]$ space whose horizontal and vertical axes correspond to "type I error" (or "false positive rate") and "1 − type II error" (or "true positive rate"), respectively. For a binary classification method, its scoring function $f(\cdot)$ estimated from the training data corresponds to an ROC curve, with every point on the curve having the coordinates (type I error, 1 − type II error) for a given threshold $c$. The area under the ROC curve (AUC) is a widely used metric to evaluate a classification method and compare different methods. In practice, the typical construction of empirical ROC curves includes the following three approaches: by varying the threshold value $c$, points on empirical ROC curves have horizontal and vertical coordinates, respectively, defined as

(approach 1) $\hat{R}_0(\phi^c)$ and $1 - \hat{R}_1(\phi^c)$ on the training data;

(approach 2) $\tilde{R}_0(\phi^c)$ and $1 - \tilde{R}_1(\phi^c)$ on the test data;

(approach 3) empirical type I error and (1 − empirical type II error) estimated by cross-validation on the training data.

Since an empirical ROC curve constructed by any of the above three approaches is an estimate of the unobserved oracle ROC curve, there is literature on the construction of confidence bands of the oracle ROC curve. For point-wise confidence intervals of points on the oracle ROC curve, typical construction methods use a binomial distribution (*34*) or a binormal distribution (*35*) to model the distribution of the given point's corresponding points on empirical ROC curves. About constructing simultaneous confidence intervals to form a confidence band of the oracle ROC curve, there are methods based on bootstrapping (*36*) or the Working-Hotelling method (*35*), among others. For a review of existing methods and an empirical study that compares them, please see (*37*). Although these construction methods of empirical ROC curves and confidence bands have many useful applications, none of them is appropriate for evaluating NP classification methods, because the empirical ROC curves and confidence bands do not provide users with direct information to find the classifiers that satisfy their pre-specified type I error upper bound $\alpha$ with high probability. Simulation S1 in the Supplementary Materials demonstrates this issue (Fig. S1).

## NP-ROC Bands

Motivated by the NP umbrella algorithm, we propose the Neyman-Pearson receiver operating characteristic (NP-ROC) bands, a second main contribution of this paper, to serve as a new visualization tool for classification methods under the NP paradigm. In the NP-ROC space, the horizontal axis is defined as the type I error upper bound (with high probability), and the vertical axis represents (1 − conditional type II error), where we define the *conditional type II error* of a classifier as its type II error conditioning on training data. An NP classifier corresponds to a vertical line segment (i.e., a blue dashed line segment in Fig. 3A) in the NP-ROC space. The horizontal coordinate of a line segment represents the type I error upper bound of that classifier. The vertical coordinates of the upper and lower ends of the segment represent the upper and lower high probability bounds of (1 − conditional type II error) of that classifier.

To create NP-ROC bands, the sample splitting scheme is slightly different from that of the umbrella algorithm. We still follow the umbrella algorithm to divide the class 0 data into two halves, using the first half to train the scoring function and the second half (size $n$) to estimate the score threshold. However, to calculate the high probability (1 − conditional type II error) bounds, we also need to divide the class 1 data into two halves, using the first half to train the scoring function and the second half to calculate the bounds. Therefore, in the construction of NP- ROC bands, we refer to the class 0 data and the first half of the class 1 data as the training data. After we train a scoring function $f$ and apply it to the left-out class 0 data, we obtain $n$ score thresholds and sort them in an increasing order. For the classifier corresponding to the $k$-th ordered score threshold, i.e., $\hat{\phi}_k$, given a pre-defined tolerance level $\delta$, we find the $(1-\delta)$ probability upper bound of $R_0(\hat{\phi}_k)$ as

$$\alpha(\hat{\phi}_k) = \inf\left\{\alpha \in [0,1]: \sum_{j=k}^{n} \binom{n}{j} (1-\alpha)^j \alpha^{n-j} \leq \delta\right\}, \tag{3}$$

because we have $\mathbb{P}[R_0(\hat{\phi}_k) \leq \alpha(\hat{\phi}_k)] \geq 1 - \sum_{j=k}^{n} \binom{n}{j} \left(1-\alpha(\hat{\phi}_k)\right)^j \left(\alpha(\hat{\phi}_k)\right)^{n-j} \geq 1-\delta$, where the first inequality follows from Equation (1). We next derive the $(1-\delta)$ high probability lower and upper bounds of the conditional type II error, denoted by $R_1^c(\hat{\phi}_k)$, as $\beta_L(\hat{\phi}_k)$ and $\beta_U(\hat{\phi}_k)$ based on Equations (S9) and (S10) in the Supplementary Materials. For every rank $k \in \{1, \cdots, n\}$, we calculate $\left(\alpha(\hat{\phi}_k), 1-\beta_U(\hat{\phi}_k)\right)$ and $\left(\alpha(\hat{\phi}_k), 1-\beta_L(\hat{\phi}_k)\right)$ for the classifier $\hat{\phi}_k$ (shown as the lower and upper ends of a blue dashed vertical line segment in Fig. 3A). Varying $k$ from 1 to $n$, we obtain n vertical line segments in the NP-ROC space. For a classifier $\hat{\phi}$ with a score threshold between two ranks of the left-out class 0 scores, say the $(k-1)$-th and the $k$-th, we have $R_0(\hat{\phi}_k) \leq R_0(\hat{\phi}) \leq R_0(\hat{\phi}_{k-1})$ and $R_1^c(\hat{\phi}_{k-1}) \leq R_1^c(\hat{\phi}) \leq R_1^c(\hat{\phi}_k)$. Hence, we set $\beta_L(\hat{\phi}) = \beta_L(\hat{\phi}_{k-1})$ and $\beta_U(\hat{\phi}) = \beta_U(\hat{\phi}_k)$. As $k$ increases from 1 to $n$, the vertical line segment shifts from right to left, and we interpolate the $n$ upper ends of these segments using right-continuous step functions, and the $n$ lower ends using left-continuous step functions. The band created after the interpolation is called an NP-ROC band (between the two black stepwise curves in Fig. 3A). This band has the interpretation that every type I error upper bound $\alpha$ corresponds to a vertical line segment, and the achievable (1− conditional type II error) is sandwiched between the lower and upper ends of the line segment with probability of at least $1-2\delta$. When we randomly split the training data for $M > 1$ times and repeat the above procedure, we obtain $M$ NP-ROC bands. For the $M$ upper curves and $M$ lower curves respectively, we calculate the average of the vertical values for every horizontal value to obtain an average upper curve and an average lower curve, which form an NP-ROC band for multiple random splits.

## Applications of NP-ROC Bands

By definition, the NP-ROC bands work naturally as a visualization tool for NP classification methods. In practice, two issues remain to be addressed in the implementation classification methods under the NP paradigm: (I) How to choose a reasonable type I error upper bound $\alpha$ if users do not have a pre-specified value in mind? (II) How to compare two classification methods under the NP paradigm? Below we use simulation and real data applications to demonstrate that the NP-ROC bands serve as an effective tool to address these two questions.

**Simulation 2.** We consider two independent predictors $X_1$ and $X_2$ with $(X_1|Y=0) \sim N(0,1)$, $(X_1|Y=1) \sim N(1,1)$, $(X_2|Y=0) \sim N(0,1)$, $(X_2|Y=1) \sim N(1,6)$, and $\mathbb{P}(Y=0) = \mathbb{P}(Y=1) = 0.5$. We simulate a data set with sample size $N = 1000$ and set the number of random splits to $M = 11$ and the tolerance level to $\delta = 0.1$. We use the Linear Discriminant Analysis (LDA) with only $X_1$ (referred to as method 1) and only $X_2$ (referred to as method 2). For each classification method, we generate its corresponding NP-ROC bands for comparison, as shown in Fig. 3B. At the horizontal axis, we mark in black the $\alpha$ values for which method 1's lower curve is higher than method 2's upper curve, and similarly we mark in red the $\alpha$ values for which method 2's lower curve is higher than method 1's upper curve. Given a tolerance level $\delta$,

users can read from these colored regions and easily decide which method performs better at any $\alpha$ values under the NP paradigm. Specifically, if users prefer a small $\alpha$ under 0.05, method 2 would be the choice between the two methods. In a different situation, suppose a user wants choose $\alpha$ for method 1 and he would like to have type II error no greater than 0.5, the NP-ROC band suggests that a reasonable 90% probability upper bound on the type I error should be greater than 0.2.

**Real Data Application 1.** We apply NP classification algorithms to data from the Early Warning Project, which has collected historical data over 50 years around the world and aims to produce risk assessments of the potential for mass atrocities. We use the public data available on the project website (*25*). We consider as the response the binary variable `mkl.start.1`, which denotes the onset of state-led mass killing episode in next year, and we use as predictors 32 variables, which are summarized in Table S1 in the Supplementary Materials. Formulating the prediction of (1 − `mkl.start.1`), for which a value 0 indicates mass killing and a value 1 indicates otherwise, as a binary classification problem, our goal is to control the type I error (the conditional probability of misclassifying a future mass killing) while optimizing the type II error. After data processing and filtering, there are 6365 observations, among which only 60 have responses as 0's. This is a scenario with extremely unbalanced classes, where the more important class 0 is the rare class. If we use the classical classification paradigm to minimize the risk, because of the dominating marginal probability of class 1, the resulting classifier would prioritize the type II error and possibly result in a large type I error, which is unacceptable in this application. On the other hand, the NP classification paradigm is specifically designed to control the type I error under a pre-specified level with high probability. We apply three NP classification methods (RF, penLR, and SVM) to this data and summarize the resulting NP-ROC bands in Fig. 4A. As a comparison, we also use 5-fold cross-validation to calculate empirical ROC curves of each method and plot the vertical average curve (for each horizontal value, we calculate the average of the 5 vertical values in the 5 empirical ROC curves), which we denote by ROC-CV, in Fig. 4B. In this application, both NP-ROC bands and ROC-CV curves indicate that RF is the best among the three methods. For direct visual comparison, we plot the NP-ROC band and the ROC-CV curve in one panel for RF (Fig. 4C) and SVM (Fig. 4D) respectively. Here we emphasize again that NP-ROC bands and ROC-CV curves have different horizontal and vertical axes. From Fig. 4C-D, it is obvious that ROC-CV curves do not provide guidance on how to determine $\alpha$ in a data adaptive way, because the horizontal axis of ROC-CV represents the empirical type I error, while NP-ROC bands have $\alpha$ as the horizontal axis and allow users to decide a reasonable $\alpha$ based on the corresponding $(1-\delta)$ high probability lower and upper bounds of the conditional type II errors. After RF is chosen as the classification method, the NP-ROC band in Fig. 4C suggests that $\alpha = 0.2$ might be a reasonable 90% probability upper bound on the type I error if political scientists desire to have the conditional type II error no greater than 0.4. We note that although the ROC-CV curve in Fig. 4C suggests that the classifier with the empirical type I error = 0.2 might be a reasonable choice, because the point on the ROC-CV curve corresponding to that classifier has the smallest distance to the point (0,1), this conclusion is made from a perspective different from the type I error control, and the chosen classifier is also different from the one corresponding to $\alpha = 0.2$.

**Real Data Application 2.** We also implement the NP umbrella algorithm and NP-ROC bands on a neuroblastoma data set of 43,827 gene expression measurements from Illumina RNA sequencing of 498 neuroblastoma samples (Gene Expression Omnibus Accession number GSE62564) generated by the Sequencing Quality Control (SEQC) Consortium (*38–41*). The gene expression measurements are available at `GSE62564_SEQC_NB_RNA-Seq_log2RPM.txt.gz` available at (*42*). Each neuroblastoma sample is labelled as high risk (HR) or non-HR, indicating whether or not the sample belongs to a high-risk patient based on clinical evidence. It is commonly understood that misclassifying an HR sample as non-HR will have more severe consequences than the other way around. Formulating this problem under the NP classification framework, we denote the HR samples as class 0 and the non-HR samples as class 1 (the classes have sample sizes 176 and 322, respectively) and use the 43,827 gene expression measurements as features to classify the samples. Setting the tolerance level as $\delta = 0.1$, we create NP-ROC bands for three classification methods: penLR, RF and NB. In Fig. S2A in Supplementary Materials, we compare penLR and RF. At every $\alpha$ value, since neither band dominates the other, we cannot distinguish between these two methods with confidence across the whole domain. In Fig. S2B in Supplementary Materials, we compare penLR and NB. The long black bar at the bottom of the plot indicates that penLR dominates NB for most of the type I upper bound $\alpha$ values. In this application, if we choose penLR or RF, Fig. S2A shows that it is reasonable to set $\alpha = 0.1$. Given $\alpha = 0.1$ and $\delta = 0.1$, after randomly splitting the data into training data with a size 374 (3/4 of the observations) and test data with a size 124 (1/4 of the observations) for 100 times, we also calculate the empirical type I and type II errors on the test data (Table S2 in Supplementary Materials). Although we can never observe the distribution of population type I and type II errors, the results show that the NP approach in practice also effectively controls the empirical type I errors under $\alpha$ with high probability, by paying the price of having larger empirical type II errors.

Besides the above three examples, in Simulation S1 in Supplementary Materials we further demonstrate that users cannot simply determine a classifier from an empirical ROC curve to control the type I error under $\alpha$ with high probability, while our proposed NP-ROC bands provide direct information for users to make such a decision. We also show in Simulation S1 and prove in Supplementary Materials that the NP-ROC lower curve is a conservative point-wise estimate of the oracle ROC curve.

## Software

The `R` package `nproc` is available at https://CRAN.R-project.org/package=nproc. The Shiny version with graphical user interface is available online at https://yangfeng.shinyapps.io/nproc/ and also available for use on local computers after users install the `R` package.
The `R` codes for generating figures and tables are available in a supporting .zip file.

## Discussion

In this paper, we propose an NP umbrella algorithm to implement scoring-type classification methods under the NP paradigm. This algorithm guarantees the desired high probability control of type I error, allowing us to construct NP classifiers in a wide range of application contexts. We also propose NP-ROC bands, a new variant of the ROC curves under the NP paradigm. NP-ROC bands provide direct information on the population type I error bounds, $\alpha$, and a range of achievable type II errors for any given $\alpha$.

We note that $n_{\min} = \lceil \log \delta \,/\, \log(1-\alpha) \rceil$ is the minimum left-out class 0 sample size to ensure that the algorithm outputs a classifier with the type I error below $\alpha$ with probability at least $(1-\delta)$. Both the type I error upper bound $\alpha$ and the tolerance level $\delta$ are subject to user's preferences. In practice, if a user has a small class 0 sample size but would like to achieve a type I error control with high probability using our umbrella algorithm, he or she must either increase $\alpha$ or $\delta$ such that the left-out class 0 sample size is above the corresponding $n_{\min}$. For commonly used $\alpha$ and $\delta$ values, $n_{\min}$ is satisfied in most applications. For example, $n_{\min} = 59$ when $\alpha = \delta = 0.05$; $n_{\min} = 45$ when $\alpha = 0.1$ and $\delta = 0.05$; $n_{\min} = 29$ when $\alpha = 0.05$ and $\delta = 0.1$. Having large class 1 and class 0 samples to train the scoring function will reduce the type II error, but it will not affect the type I error control.

One limitation of the current umbrella algorithm is the requirement of sample homogeneity, i.e., the data points in each class are independently and identically distributed. For future studies, we will generalize the umbrella algorithm for dependent data and investigate the optimality of the sample splitting ratio used in the umbrella algorithm under specific model settings.

Please note that optimality result cannot be established for the type II error without assumptions on the data distributions. In contrast to our previous work (*9*, *16*), the current work does not aim to construct an optimal classifier by estimating the two-class density ratio, which has the optimality guaranteed by the Neyman-Pearson Lemma for hypothesis testing. The reason is that the plug-in density ratio approach has difficulty with density estimation in high dimensions without restrictive assumptions, and in practice, multiple machine learning methods have been widely applied to binary classification problems with high-dimensional features. Therefore, in our current work, we developed the umbrella algorithm to integrate popular binary classification methods into the NP paradigm. Given each method, its trained scoring function and a left-out class 0 sample, our umbrella algorithm outputs a classifier that satisfies the type I error control and has the minimum type II error. In practice, users can use the NP umbrella algorithm with different classification methods and choose the method that dominate other methods at a specific $\alpha$ value or in a range of $\alpha$ values based on NP-ROC bands. Or if NP-ROC bands do not suggest clear dominance of a classification method at the users' choice of $\alpha$, users can choose the method that gives the lowest type II error by cross-validation.

It is possible to extend the NP umbrella algorithm to the multi-class classification problem. Here, we describe a simple but common scenario where users have three classes in a decreasing order of priority (e.g., class 0 = cancer of a more dangerous subtype, class 1 = cancer of a less dangerous subtype, and class 2 = benign tumor) and would like to first control the error of misclassifying class 0 and then the error of misclassifying class 1. In this scenario, we can adopt our umbrella algorithm to address the needs. A recent work provided evidence that breaking a multiple classification problem into multiple binary classification problem may lead to poor performance (*43*). One advantage of our umbrella algorithm is that it can be used with any good multi-class classification algorithms, and it does not need to decompose the classification problem into multiple binary classification problems if the algorithm does not do so. The basic idea of using the umbrella algorithm in the multi-class case is as follows. Given a trained multi-class classification algorithm, we first apply it to the left-out class 0 data to obtain the probabilities of assigning them to class 0. With a pre-specified $(1-\delta)$ probability upper bound $\alpha_0$ on the error $\mathbb{P}(\phi(X) \neq Y | Y = 0)$, where $\phi(X)$ denotes the predicted class label, and $Y$ denotes the actual class label, we can use Proposition 1 to find a threshold $c_0$ for assigning a new data point to class 0. Similarly, we can apply the trained algorithm to the left-out class 1 data to obtain the probabilities of assigning them to class 1. With a pre-specified $(1-\delta)$ probability upper bound $\alpha_1$ on the error $\mathbb{P}(\phi(X) \neq Y | Y = 1)$, we can use Proposition 1 to find another threshold $c_1$ for assigning a new data point to class 1. With these two thresholds, we can construct the following classifier: given a new data point, we use the trained algorithm to estimate its probabilities of being assigned to classes 0 and 1, denoted as $\hat{p}_0$ and $\hat{p}_1$ respectively. If $\hat{p}_0 \geq c_0$, we will assign this new data point to class 0. If $\hat{p}_0 < c_0$ and $\hat{p}_1 \geq c_1$, we will assign the new data point to class 1. Otherwise, we will assign it to class 2.

## Acknowledgments

**Funding:** This work was supported by Zumberge Individual Research Award (X.T.), NSF grants DMS-1613338 (X.T. and J.J.L.), DMS-1308566 (Y.F.), DMS-1554804 (Y.F.), Hellman Fellowship (J.J.L.), NIH grant R01GM120507 (J.J.L. and X.T.), and the PhRMA Foundation Research Starter Grant in Informatics (J.J.L.).
**Author contributions:** All the authors designed and conducted the research, discussed the results and contributed to the manuscript writing.
**Competing interests:** The authors declare that they have no competing interests.

**Figures**

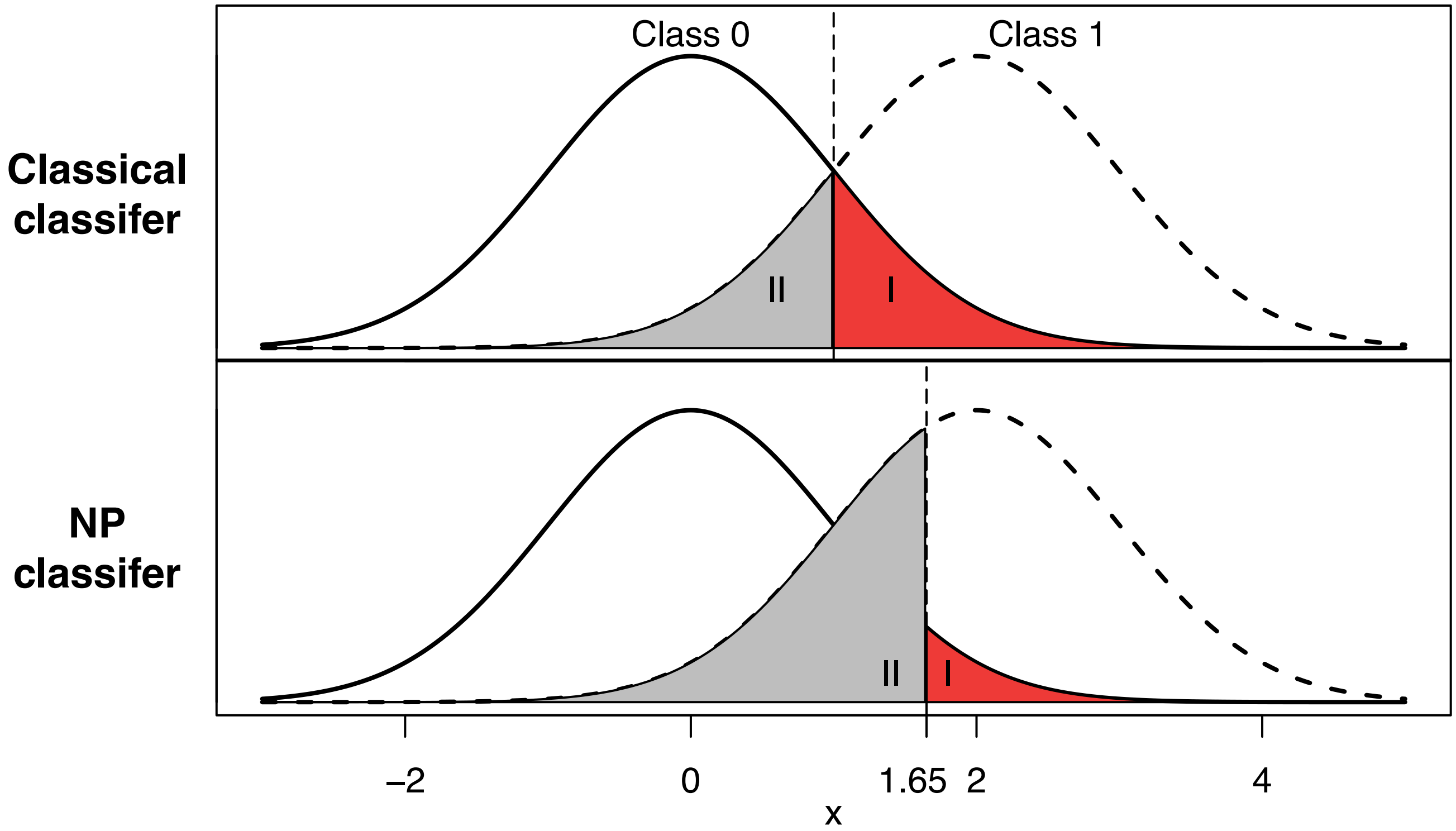

**Fig. 1. Classical vs. NP oracle classifiers in a binary classification example.** In this toy example, the two classes have equal marginal probabilities, i.e., $\mathbb{P}(Y = 0) = \mathbb{P}(Y = 1) = 0.5$. Suppose that a user prefers a type I error $\leq 0.05$. The classical classifier $I(X > 1)$ that minimizes the risk would result in a type I error = 0.16. On the other hand, the NP classifier $I(X > 1.65)$ that minimizes the type II error under the type I error constraint ($\alpha = 0.05$) delivers a desirable type I error. This figure is adapted from (*17*, *44*) with permissions.

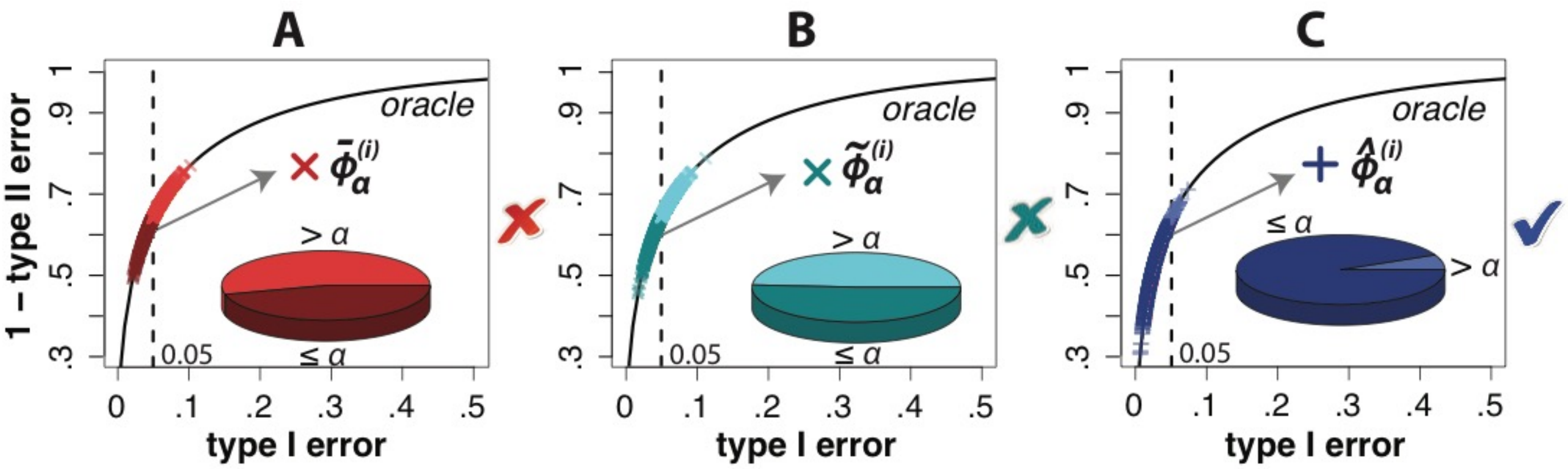

**Fig. 2. Choose a threshold such that the type I error is below $\alpha$ with high probability.** (**A**) The naïve approach. (**B**) The 5-fold cross-validation based approach. (**C**) The NP approach.

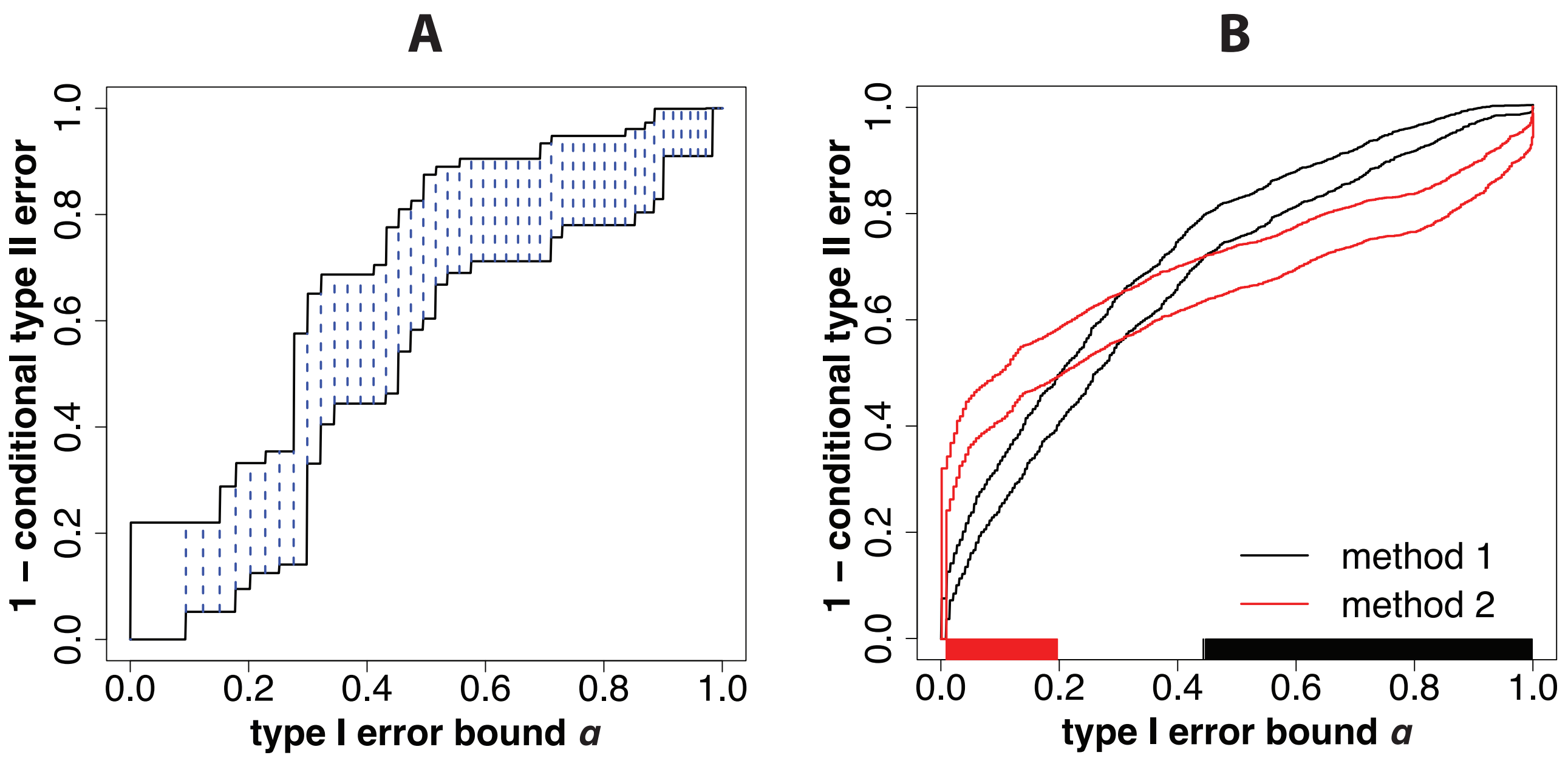

**Fig. 3. Illustration of NP-ROC bands.** (**A**) How to draw an NP-ROC band. Each blue dashed line represents one NP classifier, with horizontal coordinate $\alpha$ and vertical coordinates $1 - \beta_U$ (lower) and $1 - \beta_L$ (upper). Right-continuous and left-continuous step functions are used to interpolate points on the upper and lower ends, respectively. (**B**) Use of NP-ROC bands to compare the two LDA methods in Simulation 2.

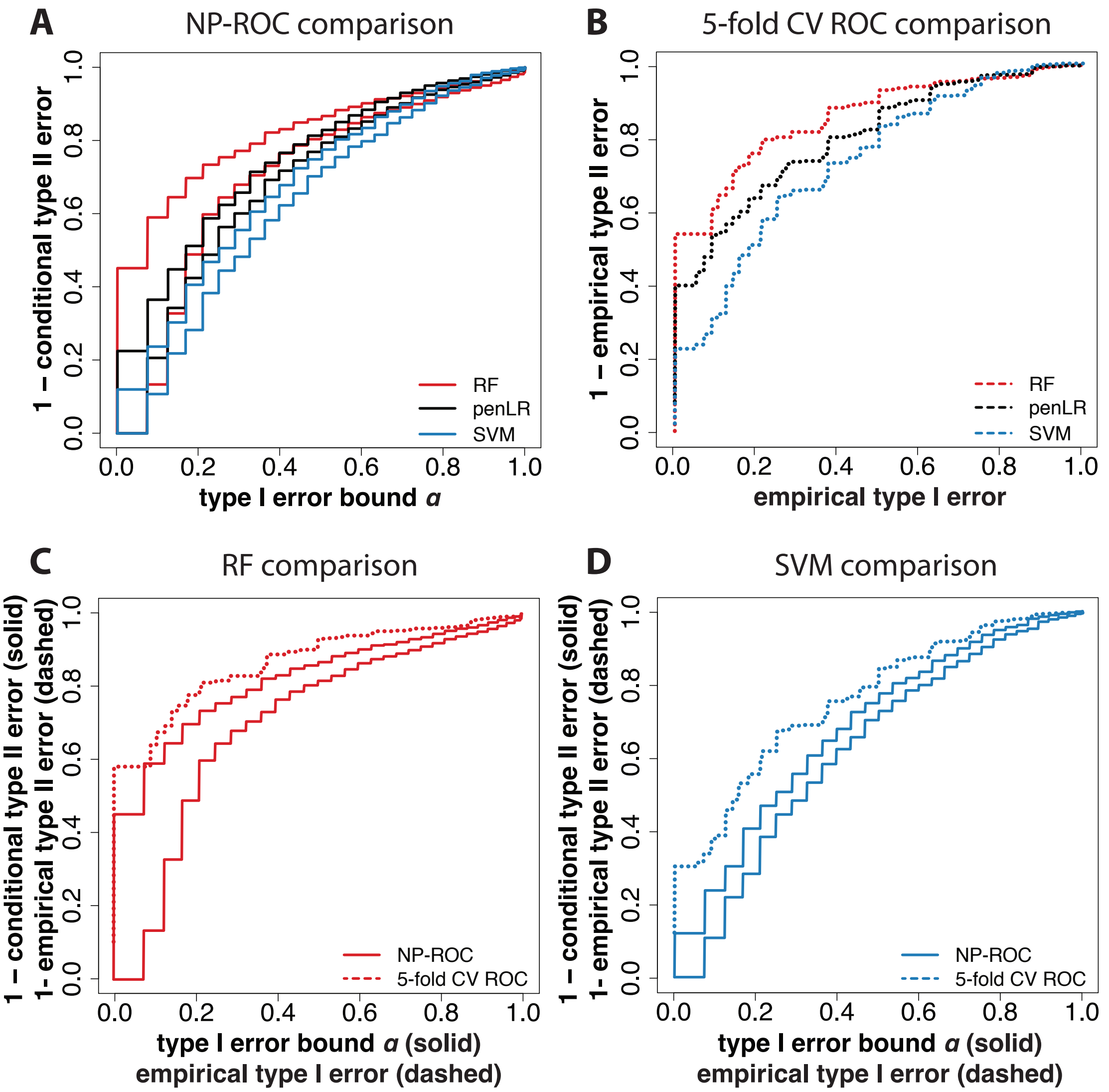

**Fig. 4. NP-ROC bands and ROC-CV curves (generated by 5-fold cross-validation) of three classification methods in real data application 1.** (**A**) NP-ROC bands of random forests (RF) vs. penalized logistic regression (penLR) vs. support vector machines (SVM). RF dominates the other two methods for a wide range of α values. (**B**) ROC-CV curves of RF, penLR and SVM. Among the three methods, RF has the largest area under the curve. (**C**) NP-ROC band and ROC-CV curve of RF. (**D**) NP-ROC band and ROC-CV curve of SVM.

## Supplementary Materials

### Proof of Proposition 1

**Proof:** In our context, let $T_{(k)}$ be the $k$-th ordered classification score of a left-out class 0 sample (i.e., class 0 sample not used to train a base algorithm). Suppose $T_{(k)}$ is the chosen threshold on classification scores. Then the classifier is $\hat{\phi}_k = (T > T_{(k)})$, where $T$ is the classification score of an independent observation from class 0, and the population type I error of $\hat{\phi}_k$ given $T_{(k)}$ is

$$\mathbb{P}[T > T_{(k)} | T_{(k)}] = 1 - F(T_{(k)}),$$

where $F$ is the cumulative distribution function of $T_i$'s.

Hence, the violation rate (the probability that the population type I error of $\hat{\phi}_k$ exceeds $\alpha$) is

$$\begin{aligned}
\mathbb{P}\left[1 - F\left(T_{(k)}\right) > \alpha\right] &= \mathbb{P}\left[F\left(T_{(k)}\right) < 1 - \alpha\right] = \mathbb{P}\left[T_{(k)} < F^{-1}(1-\alpha)\right] \\
&= \mathbb{P}\left[T_{(1)} < F^{-1}(1-\alpha), \cdots, T_{(k)} < F^{-1}(1-\alpha)\right] \\
&= \mathbb{P}[\text{at least } k \text{ of the } T_i\text{'s are less than } F^{-1}(1-\alpha)] \\
&= \sum_{j=k}^{n} \mathbb{P}[\text{exactly } j \text{ of the } T_i\text{'s are less than } F^{-1}(1-\alpha)] \\
&= \sum_{j=k}^{n} \binom{n}{j} \mathbb{P}[T_i < F^{-1}(1-\alpha)]^j (1 - \mathbb{P}[T_i < F^{-1}(1-\alpha)])^{n-j} \\
&\leq \sum_{j=k}^{n} \binom{n}{j} (1-\alpha)^j \alpha^{n-j}, \qquad \text{(S1)}
\end{aligned}$$

where the last inequality holds because $\mathbb{P}[T_i < F^{-1}(1-\alpha)] \leq 1 - \alpha$, and it becomes an equality when $F$ is continuous.

**Remark:** The proof does not have any assumptions on $F$. Regardless of the continuity of $F$, we define its inverse as $F^{-1}(\cdot) = \inf\{x: F(x) \leq \cdot\}$, which has the property: $x \leq F(y)$ if and only if $F^{-1}(x) \leq y$, for any $x \in [0,1]$ and $y$ in the domain of $F$.

**Conditional type II error bounds in NP-ROC Bands**

Given training data $\mathcal{S} = \mathcal{S}^0 \cup \mathcal{S}^1$, where $\mathcal{S}^0$ and $\mathcal{S}^1$ are class 0 and class 1 samples, respectively, we randomly split $\mathcal{S}^0$ into $\mathcal{S}_1^0$ and $\mathcal{S}_2^0$ and split $\mathcal{S}^1$ into $\mathcal{S}_1^1$ and $\mathcal{S}_2^1$. For simplicity, we let $|\mathcal{S}_1^0| = |\mathcal{S}_2^0| = n$ and $|\mathcal{S}_1^1| = |\mathcal{S}_2^1| = m$, and express the two left-out samples as $\mathcal{S}_2^0 = \{x_1^0, \cdots, x_n^0\}$ and $\mathcal{S}_2^1 = \{X_1^1, \cdots, X_m^1\}$. In the following discussion, we treat $\mathcal{S}_1^0$, $\mathcal{S}_2^0$ and $\mathcal{S}_1^1$ as fixed (by conditioning on them) and only consider the $m$ data points in $\mathcal{S}_2^1$ as random variables.

We train a base classification algorithm (e.g., SVM) on $\mathcal{S}_1^0 \cup \mathcal{S}_1^1$ and denote the resulting classification scoring function as $f$. Because $f$ is a function of $\mathcal{S}_1^0 \cup \mathcal{S}_1^1$, in our discussion here, $f$ is a fixed function that maps $\mathcal{X}$ to $\mathbb{R}$.

After applying $f$ to the left-out samples $\mathcal{S}_2^0$ and $\mathcal{S}_2^1$, we denote the resulting classification scores as $t_i^0 = f(x_i^0), i = 1, \cdots, n$, and $T_j^1 = f(X_j^1), j = 1, \cdots, m$, respectively.

Suppose that we decide to use the $k$-th ordered left-out class 0 score, $t_{(k)}^0$, as the score threshold. We then construct an NP classifier $\hat{\phi}_k$ (based on one ($M = 1$) random split) as

$$\hat{\phi}_k(X) = I\left(f(X) > t_{(k)}^0\right).$$

We then find the corresponding rank of $t_{(k)}^0$ among the left-out class 1 scores $T_1^1, \cdots, T_m^1$.

There are three scenarios.

**Scenario 1**

If $T_{(1)}^1 \le t_{(k)}^0 \le T_{(m)}^1$, we define the lower bound rank $r_L$ and the upper bound rank $r_U$ as

$$r_L = \max\left\{r \in \{1, \cdots, m\}: T_{(r)}^1 \le t_{(k)}^0\right\}, \tag{S2}$$

$$r_U = \min\left\{r \in \{1, \cdots, m\}: T_{(r)}^1 \ge t_{(k)}^0\right\}, \tag{S3}$$

and denote their corresponding classifiers as

$$\tilde{\phi}_{r_L}(X) = I\left(f(X) > T_{(r_L)}^1\right), \tag{S4}$$

$$\tilde{\phi}_{r_U}(X) = I\left(f(X) > T_{(r_U)}^1\right). \tag{S5}$$

We define the conditional (here the conditioning is on training data $\mathcal{S}_1^0$, $\mathcal{S}_2^0$ and $\mathcal{S}_1^1$) type II errors of $\tilde{\phi}_{r_L}$ and $\tilde{\phi}_{r_U}$ as

$$R_1^c\left(\tilde{\phi}_{r_L}\right) := \mathbb{P}\left[f(X^1) \le T_{(r_L)}^1 | T_{(r_L)}^1\right] = F_1\left(T_{(r_L)}^1\right), \tag{S6}$$

$$R_1^c\left(\tilde{\phi}_{r_U}\right) := \mathbb{P}\left[f(X^1) \le T_{(r_U)}^1 | T_{(r_U)}^1\right] = F_1\left(T_{(r_U)}^1\right), \tag{S7}$$

where $X^1$ represents a new class 1 observation, and $F_1$ is the cumulative distribution function of classification scores of class 1 observations.

Because $T_{(r_L)}^1 \le t_{(k)}^0 \le T_{(r_U)}^1$, we have

$$R_1^c\left(\tilde{\phi}_{r_L}\right) \le R_1^c\left(\hat{\phi}_k\right) \le R_1^c\left(\tilde{\phi}_{r_U}\right), \tag{S8}$$

where $R_1^c(\cdot)$ stands for the conditional type II error, and $R_1^c\left(\hat{\phi}_k\right) = \mathbb{P}\left[f(X^1) \le t_{(k)}^0\right] = F_1\left(t_{(k)}^0\right)$. In (S8), the left inequality becomes tight when $T_{(r_L)}^1 = t_{(k)}^0$ and the right inequality is tight when $T_{(r_U)}^1 = t_{(k)}^0$.

Given a pre-specified tolerance level $\delta$, denote by $\beta_L(\hat{\phi}_k)$ and $\beta_U(\hat{\phi}_k)$ the $(1-\delta)$ high probability lower and upper bounds of $R_1^c\left(\hat{\phi}_k\right)$. Specifically, $\beta_L(\hat{\phi}_k)$ and $\beta_U\left(\hat{\phi}_k\right)$ are defined as

$$\beta_L(\hat{\phi}_k) := \sup\left\{\beta \in [0,1]: \sum_{j=r_L}^{m} \binom{m}{j} \beta^j (1-\beta)^{m-j} \le \delta\right\}, \tag{S9}$$

$$\beta_U\left(\hat{\phi}_k\right) := \inf\left\{\beta \in [0,1]: \sum_{j=r_U}^{m} \binom{m}{j} \beta^j (1-\beta)^{m-j} \ge 1-\delta\right\}. \tag{S10}$$

The reason that (S9) and (S10) give valid $(1-\delta)$ high probability lower and upper bounds is as follows.

- A constant $\beta$ serves as a valid $(1-\delta)$ high probability lower bound of $R_1^c(\hat{\phi}_k)$ if
$$\mathbb{P}\left[R_1^c(\hat{\phi}_k) \geq \beta\right] \geq 1-\delta\,. \tag{S11}$$
Since by (S8),
$$\mathbb{P}\left[R_1^c(\hat{\phi}_k) \geq \beta\right] \geq \mathbb{P}\left[R_1^c(\tilde{\phi}_{r_L}) \geq \beta\right],$$
in order to have (S11) hold it is sufficient to have
$$\mathbb{P}\left[R_1^c(\tilde{\phi}_{r_L}) \geq \beta\right] \geq 1-\delta\,.$$
By (S6),
$$\begin{aligned}\mathbb{P}\left[R_1^c(\tilde{\phi}_{r_L}) \geq \beta\right] &= \mathbb{P}\left[F_1\left(T^1_{(r_L)}\right) \geq \beta\right] = 1-\mathbb{P}\left[F_1\left(T^1_{(r_L)}\right) < \beta\right] \\ &\geq 1-\sum_{j=r_L}^{m}\binom{m}{j}\beta^j(1-\beta)^{m-j},\end{aligned}$$
where the inequality in the second line follows from (S1) in the proof of Proposition 1. Hence, it suffices to have
$$\sum_{j=r_L}^{m}\binom{m}{j}\beta^j(1-\beta)^{m-j} \leq \delta$$
to make (S11) hold. Among all the $\beta$ values that satisfy (S11), we would choose the supremum as the $(1-\delta)$ high probability lower bound of $R_1^c(\hat{\phi}_k)$, leading to (S9).
- A constant $\beta$ serves as a valid $(1-\delta)$ high probability upper bound of $R_1^c(\hat{\phi}_k)$ if
$$\mathbb{P}\left[R_1^c(\hat{\phi}_k) \leq \beta\right] \geq 1-\delta\,. \tag{S12}$$
Since by (S8),
$$\mathbb{P}\left[R_1^c(\hat{\phi}_k) \leq \beta\right] \geq \mathbb{P}\left[R_1^c(\tilde{\phi}_{r_U}) \leq \beta\right],$$
in order to have (S12) hold it is sufficient to have
$$\mathbb{P}\left[R_1^c(\tilde{\phi}_{r_U}) \leq \beta\right] \geq 1-\delta\,.$$
By (S7),
$$\begin{aligned}\mathbb{P}\left[R_1^c(\tilde{\phi}_{r_U}) \leq \beta\right] &= \mathbb{P}\left[F_1\left(T^1_{(r_U)}\right) \leq \beta\right] \\ &= \mathbb{P}\left[F_1\left(T^1_{(r_U)}\right) < \beta\right] = \sum_{j=r_U}^{m}\binom{m}{j}\beta^j(1-\beta)^{m-j},\end{aligned}$$
where the two equalities in the second line follow from (S1) in the proof of Proposition 1, under the assumption that $F_1$ is continuous, which holds for most classification algorithms. Hence, it suffices to have
$$\sum_{j=r_U}^{m}\binom{m}{j}\beta^j(1-\beta)^{m-j} \geq 1-\delta$$
to make (S12) hold. Among all the $\beta$ values that satisfy (S12), we would choose the infimum as the $(1-\delta)$ high probability upper bound of $R_1^c(\hat{\phi}_k)$, leading to (S10).

**Scenario 2**

If $t^0_{(k)} > T^1_{(m)}$, we define the lower bound rank $r_L$ the same as in (S2) and the $(1-\delta)$ high probability lower bound $\beta_L(\hat{\phi}_k)$ the same as in (S9). We set the $(1-\delta)$ high probability upper bound $\beta_U(\hat{\phi}_k) = 1$.

**Scenario 3**

If $t_{(k)}^{0} < T_{(1)}^{1}$, we define the upper bound rank $r_U$ the same as in (S3) and the $(1-\delta)$ high probability upper bound $\beta_U(\hat{\phi}_k)$ the same as in (S10). We set the $(1-\delta)$ high probability lower bound $\beta_L(\hat{\phi}_k) = 0$.

In all the above three scenarios, we have $\mathbb{P}\left[R_1^c(\hat{\phi}_k) < \beta_L(\hat{\phi}_k)\right] \leq \delta$ and $\mathbb{P}\left[R_1^c(\hat{\phi}_k) > \beta_U(\hat{\phi}_k)\right] \leq \delta$, leading to

$$\mathbb{P}\left[\beta_L(\hat{\phi}_k) \leq R_1^c(\hat{\phi}_k) \leq \beta_U(\hat{\phi}_k)\right] \geq 1 - 2\delta \ .$$

**Empirical ROC curves vs. NP-ROC bands in guiding users to choose classifiers to satisfy type I error control**

In practice, the popular ROC curves cannot provide proper guidance for comparing two classifiers whose type I errors are bounded from above by some $\alpha$, because ROC curves are constructed based on empirical type I and type II errors, which are calculated from test data or cross-validation on training data and do not display population type I error information. We refer to such ROC curves as empirical ROC curves in the main text to differentiate them from the oracle ROC curves. Concretely, given an empirical ROC curve of a classification method, it is unclear how users should decide which point on the curve corresponds to a classifier satisfying the type I error bound $\alpha$. Through Simulation 1 and Fig. 2, we showed that users cannot simply pick the point that has empirical type I error (i.e., horizontal axis of the ROC curve) no greater than and closest to $\alpha$, a seemingly intuitive but actually improper practice. We further illustrate this point in Simulation S1 and Fig. S1. Due to the lack of direct information on population type I errors, existing methods for constructing ROC confidence bands cannot serve the purpose either.

**Simulation S1.** From the same setup as in Simulation 1:

$$(X|Y=0) \sim N(0,1) \text{ and } (X|Y=1) \sim N(2,1), \text{ with } \mathbb{P}(Y=0)=\mathbb{P}(Y=1)=0.5\,,$$

we simulate $2D = 2000$ data sets $\left\{(x_i^{(m)}, y_i^{(m)})\right\}_{i=1}^{N}$, where $m = 1,\cdots,2D$ and $N = 1000$. The first $D$ data sets are used as the training data, and the other $D$ data sets are the test data. On the $m$-th training data set, we construct $N$ classifiers, $I(X > x_i^{(m)})$, $i = 1,\cdots,N$, and evaluate their empirical type I and II errors on the $m$-th test data set (i.e., the $(D+m)$-th simulated data set), resulting in one ROC curve. We also use the $m$-th training data set to calculate one NP-ROC lower curve, i.e., the lower curve of an NP-ROC band. Fig. S1 illustrates the $D = 1000$ ROC and NP-ROC lower curves. Suppose that users would like to find a classifier respecting a type I error bound $\alpha = 0.05$ with tolerance level $\delta = 0.05$ from an ROC curve. An intuitive choice is to pick the classifier at the intersection of the ROC curve and the vertical line at $\alpha$. If there is no classifier right at the intersection, a reasonable idea is to pick the first classifier to the left of the intersection. For the $D$ classifiers chosen in this way, we summarize their empirical type I errors (on the test data) and their population type I errors as histograms (Fig. S1 left panel). The results suggest that although the classifiers have no *empirical* type I errors greater than $\alpha$, approximately 30% of the classifiers have *population* type I errors greater than $\alpha$, violating users' desire for controlling type I error under $\alpha$ with at least 0.95 probability. On the other hand, the NP-ROC lower curves provide a natural way for users to choose classifiers given $\alpha$, as the horizontal coordinates of the NP-ROC curves are type I error upper bounds. Users can simply pick the classifier with horizontal coordinate $\alpha$. For the $D$ chosen NP classifiers, we summarize their empirical type I errors on the test data and their population type I errors as histograms (Fig. S1 right panel). It is clear that the violation rate of population type I error is under $\delta = 0.05$.

Another use of the NP-ROC lower curve is that it provides a conservative point-wise estimate of the oracle ROC curve. For an NP classifier $\hat{\phi}$, the corresponding point on an NP-ROC lower curve is, with high probability, below and to the right of the point on the oracle ROC curve. To explain this phenomenon, suppose that the classifier corresponds to the point $\left(\alpha(\hat{\phi}), 1-\beta_U(\hat{\phi})\right)$ on an NP-ROC lower curve and the point $\left(R_0(\hat{\phi}), 1-R_1(\hat{\phi})\right)$ on an oracle ROC curve. Then, by the definition of $\alpha(\hat{\phi})$ (Equation (3)) and $\beta_U(\hat{\phi})$ (Equation (S10)) as the high probability upper bounds on $R_0(\hat{\phi})$ and $R_1^c(\hat{\phi})$ (the conditional type II error, conditioning on the training data), respectively, we know that $\alpha(\hat{\phi}) \geq R_0(\hat{\phi})$ and $1-\beta_U(\hat{\phi}) \leq 1-R_1^c(\hat{\phi})$ hold with high probability. As $R_1(\hat{\phi}) = \mathbb{E}\left[R_1^c(\hat{\phi})\,\right]$, where the expectation is with respect to the joint distribution of the training data, we have $1-\beta_U(\hat{\phi}) \leq 1-R_1(\hat{\phi})$. Hence, the point on the NP-ROC lower curve is below and to the right of the oracle ROC curve with high probability. In other words, for an NP classifier, coordinates on the NP-ROC lower curve provide a conservative estimate of the corresponding coordinates on the oracle ROC curve. This phenomenon is visualized in the top right panel of Fig. S1.

**Effects of Majority Voting on the Type I and Type II Errors of the Ensemble Classifier**

In the following simulation, we demonstrate that the ensemble classifiers from multiple random splits maintain the type I error violation rates under $\delta$ and achieve reduced average type II errors with smaller standard errors over multiple simulations, as compared with the NP classifiers from just one random split.

**Simulation S2.** Our first generative model is a logistic regression (LR) model with $d = 3$ independent features and $n = 1000$ observations. For each feature, $1000$ values are independently drawn from a standard Normal distribution. The three features have non-zero coefficients as 3, 2.4 and 1.8, respectively. Denoting the feature matrix by $\mathbf{X} = [\mathbf{x}_1, \cdots, \mathbf{x}_n]^T \in \mathbb{R}^{n\times d}$ and the coefficient vector by $\beta \in \mathbb{R}^d$, we simulate the response vector as

$$\mathbf{Y} = (Y_1, \cdots, Y_n)^T, \text{ with } Y_i \overset{\text{indep}}{\sim} \text{Bernoulli}\left(\frac{1}{1+\exp(-\mathbf{x}_i^T\beta)}\right).$$

Our second generative model is a linear discriminant analysis (LDA) model with $d = 3$ independent features and $n = 1000$ observations. We first simulate the response vector as

$$\mathbf{Y} = (Y_1, \cdots, Y_n)^T, \text{ with } Y_i \overset{\text{indep}}{\sim} \text{Bernoulli}(0.5).$$

Then we simulate the feature matrix $\mathbf{X} = [\mathbf{x}_1, \cdots, \mathbf{x}_n]^T \in \mathbb{R}^{n\times d}$ as

$$(\mathbf{x}_i | Y_i = 0) \sim N((0,0,0)^T, \mathbf{I}_3);\ (\mathbf{x}_i | Y_i = 1) \sim N((2,1.6,1.2)^T, \mathbf{I}_3).$$

We simulate $1000$ datasets from each of the above two models, obtaining $2000$ datasets. With $\alpha = 0.05$, $\delta = 0.05$, and varying number of splits $M \in \{1, 5, 9, 11, 15\}$, we construct five NP classifiers for each of six methods (logistic regression, support vector machines, random forests, naïve Bayes, linear discriminant analysis, and AdaBoost) with different numbers of splits based on each data set. We separately simulate two large test data sets with $10^6$ observations from the two models, so that we can use the test data set to calculate the test (empirical) type I and type II errors to approximate the population type I and type II errors.

For each combination of $M$ value, classification method and generative model, we use its corresponding $1000$ NP classifiers to calculate average of the approximate type I errors and its standard error, the percentage of classifiers with approximate type I errors greater than $\alpha$ (type I error violation rate), and the average of the approximate type II errors and its standard error. The results are summarized below.

**Logistic Regression**

| Model | $M$ | Type I error avg (se) | Type I error violation rate | Type II error avg (se) |
|---|---|---|---|---|
| LR | 1 | 2.78% (1.01%) | 2.30% | 33.42% (5.19%) |
| | 5 | 2.72% (0.80%) | 1.00% | 33.38% (4.12%) |
| | 9 | 2.71% (0.75%) | 0.50% | 33.35% (3.89%) |
| | 11 | 2.71% (0.74%) | 0.50% | 33.37% (3.86%) |
| | 15 | 2.70% (0.73%) | 0.50% | 33.35% (3.80%) |
| LDA | 1 | 2.82% (1.06%) | 3.20% | 19.22% (4.62%) |
| | 5 | 2.74% (0.82%) | 0.70% | 19.14% (3.63%) |
| | 9 | 2.72% (0.78%) | 0.40% | 19.13% (3.39%) |
| | 11 | 2.72% (0.77%) | 0.40% | 19.09% (3.33%) |
| | 15 | 2.72% (0.75%) | 0.40% | 19.10% (3.27%) |

**Support Vector Machines**

| Model | $M$ | Type I error avg (se) | Type I error violation rate | Type II error avg (se) |
|---|---|---|---|---|
| LR | 1 | 2.81% (1.04%) | 3.00% | 43.22% (9.40%) |
| | 5 | 2.70% (0.77%) | 0.90% | 42.85% (7.53%) |
| | 9 | 2.69% (0.73%) | 0.50% | 42.76% (7.25%) |
| | 11 | 2.68% (0.72%) | 0.70% | 42.81% (7.20%) |
| | 15 | 2.68% (0.71%) | 0.50% | 42.74% (7.15%) |
| LDA | 1 | 2.81% (1.08%) | 3.40% | 34.11% (15.42%) |
| | 5 | 2.72% (0.78%) | 0.90% | 33.25% (12.95%) |
| | 9 | 2.72% (0.76%) | 0.60% | 33.05% (12.75%) |
| | 11 | 2.72% (0.75%) | 0.60% | 33.10% (12.73%) |
| | 15 | 2.71% (0.74%) | 0.50% | 33.15% (12.58%) |

**Random Forests**

| Model | $M$ | Type I error avg (se) | Type I error violation rate | Type II error avg (se) |
|---|---|---|---|---|
| LR | 1 | 2.79% (1.03%) | 2.70% | 45.08% (7.52%) |
| | 5 | 2.39% (0.72%) | 0.20% | 45.05% (5.52%) |
| | 9 | 2.32% (0.66%) | 0.00% | 44.97% (5.11%) |
| | 11 | 2.31% (0.65%) | 0.10% | 44.96% (5.04%) |
| | 15 | 2.29% (0.64%) | 0.00% | 44.93% (5.00%) |
| LDA | 1 | 2.80% (1.08%) | 3.30% | 26.00% (7.16%) |
| | 5 | 2.60% (0.75%) | 0.20% | 25.08% (4.92%) |
| | 9 | 2.58% (0.73%) | 0.40% | 24.84% (4.60%) |
| | 11 | 2.57% (0.72%) | 0.40% | 24.81% (4.52%) |
| | 15 | 2.56% (0.70%) | 0.30% | 24.77% (4.42%) |

**Naïve Bayes**

| Model | $M$ | Type I error avg (se) | Type I error violation rate | Type II error avg (se) |
|---|---|---|---|---|
| LR | 1 | 2.79% (1.01%) | 2.00% | 35.31% (5.32%) |
| | 5 | 2.69% (0.76%) | 0.50% | 35.18% (4.11%) |
| | 9 | 2.68% (0.72%) | 0.40% | 35.13% (3.87%) |
| | 11 | 2.67% (0.71%) | 0.30% | 35.15% (3.84%) |
| | 15 | 2.67% (0.70%) | 0.20% | 35.11% (3.75%) |
| LDA | 1 | 2.81% (1.05%) | 3.50% | 19.20% (4.58%) |
| | 5 | 2.73% (0.82%) | 0.80% | 19.11% (3.62%) |
| | 9 | 2.72% (0.77%) | 0.40% | 19.06% (3.35%) |
| | 11 | 2.72% (0.75%) | 0.40% | 19.03% (3.28%) |
| | 15 | 2.71% (0.75%) | 0.40% | 19.05% (3.24%) |

**Linear Discriminant Analysis**

| Model | $M$ | Type I error avg (se) | Type I error violation rate | Type II error avg (se) |
|---|---|---|---|---|
| | 1 | 2.78% (1.02%) | 3.00% | 33.48% (5.19%) |
| | 5 | 2.72% (0.80%) | 1.40% | 33.43% (4.14%) |
| LR | 9 | 2.71% (0.75%) | 0.80% | 33.42% (3.90%) |
| | 11 | 2.70% (0.74%) | 0.50% | 33.44% (3.85%) |
| | 15 | 2.70% (0.73%) | 0.30% | 33.43% (3.82%) |
| | 1 | 2.81% (1.06%) | 3.60% | 19.15% (4.59%) |
| | 5 | 2.73% (0.82%) | 0.70% | 19.11% (3.64%) |
| LDA | 9 | 2.72% (0.77%) | 0.40% | 19.05% (3.36%) |
| | 11 | 2.72% (0.76%) | 0.30% | 19.04% (3.29%) |
| | 15 | 2.71% (0.75%) | 0.40% | 19.05% (3.25%) |

**AdaBoost**

| Model | $M$ | Type I error avg (se) | Type I error violation rate | Type II error avg (se) |
|---|---|---|---|---|
| | 1 | 2.81% (1.02%) | 2.90% | 41.05% (6.20%) |
| | 5 | 2.39% (0.70%) | 0.00% | 40.87% (4.51%) |
| LR | 9 | 2.33% (0.65%) | 0.00% | 40.74% (4.19%) |
| | 11 | 2.31% (0.64%) | 0.00% | 40.77% (4.14%) |
| | 15 | 2.29% (0.62%) | 0.00% | 40.72% (4.06%) |
| | 1 | 2.80% (1.09%) | 3.60% | 24.39% (6.21%) |
| | 5 | 2.55% (0.74%) | 0.80% | 23.39% (4.20%) |
| LDA | 9 | 2.51% (0.71%) | 0.50% | 23.21% (3.93%) |
| | 11 | 2.50% (0.70%) | 0.30% | 23.17% (3.86%) |
| | 15 | 2.49% (0.68%) | 0.20% | 23.11% (3.78%) |

The results show that the type I error violation rates of the ensemble classifiers stay below $\delta = 5\%$, the average type I and type II errors, which approximate the average of the population type I and type II errors of the ensemble classifiers, and their standard deviations decrease from $M = 1$ to $M > 1$.

This simulation experiment and its results demonstrate the validity and effectiveness of the ensemble approach by majority voting.

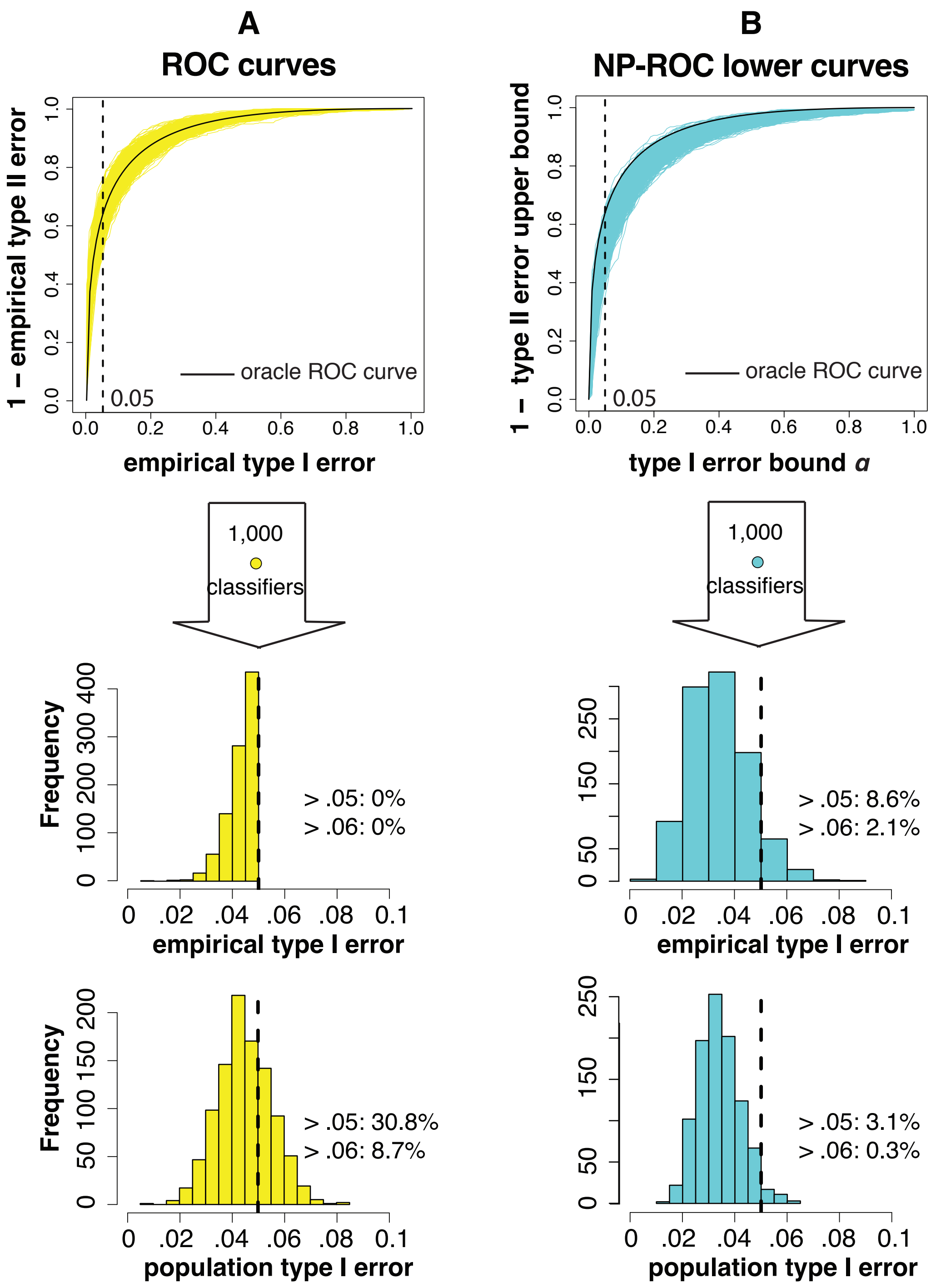

**Fig. S1. Illustration of choosing NP classifiers from empirical ROC curves and NP-ROC lower curves in Simulation S1.** (**A**) Distributions of empirical type I errors and population type I errors of 1000 classifiers, with each classifier chosen from one empirical ROC curve corresponding to the largest empirical type I error no greater than 0.05. (**B**) Distributions of empirical type I errors and population type I errors of 1000 classifiers, with each classifier chosen from one NP-ROC lower curve ($\delta = 0.05$) corresponding to $\alpha = 0.05$.

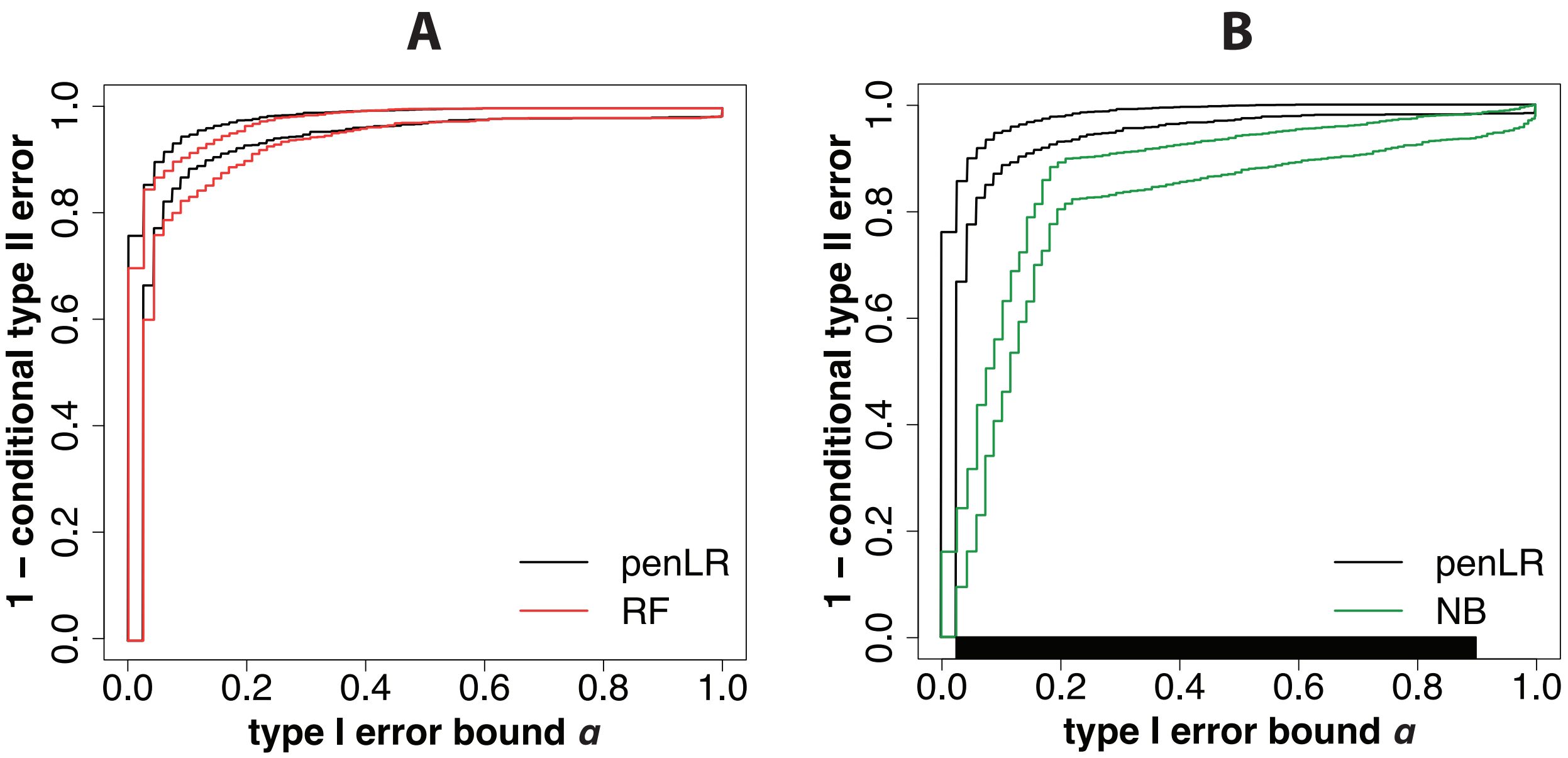

**Fig. S2. NP-ROC bands of three classification methods in real data application 2.** (**A**) Penalized logistic regression (penLR) vs. random forests (RF). No method dominates the other for any $\alpha$ values. (**B**) penLR vs. naïve Bayes (NB). The black bar at the bottom indicates the $\alpha$ values where penLR is better than NB with high probability.

| | Variable | Description | Values |
|---|---|---|---|
| Response | `mkl.start.1` | Onset of state-led mass killing episode in next year $(t+1)$ | {0, 1} |
| Predictors | `reg.afr` | US Dept State region: Sub-Saharan Africa | {0, 1} |
| | `reg.eap` | US Dept State region: East Asia and Pacific | {0, 1} |
| | `reg.eur` | US Dept State region: Europe and Eurasia | {0, 1} |
| | `reg.mna` | US Dept State region: Middle East and North Africa | {0, 1} |
| | `reg.sca` | US Dept State region: South and Central Asia | {0, 1} |
| | `reg.amr` | US Dept State region: Americas | {0, 1} |
| | `mkl.ongoing` | Any ongoing episodes of state-led mass killing | {0, 1} |
| | `mkl.ever` | Any state-led mass killing since WWII (cumulative) | {0, 1} |
| | `countryage.ln` | Country age, logged | [0, 7.712891] |
| | `wdi.popsize.ln` | Population size, logged | [4.781189, 14.130934] |
| | `imr.normed.ln` | Infant mortality rate relative to annual global median, logged | [−2.721325, 1.798977] |
| | `gdppcgrow.sr` | Annual % change in GDP per capita, meld of IMF and WDI, square root | [−8.002944, 13.777878] |
| | `wdi.trade.ln` | Trade openness, logged | [−3.863269, 6.276150] |
| | `ios.iccpr1` | ICCPR 1st Optional Protocol signatory | {0, 1} |
| | `postcw` | Post-Cold War period (year ≥ 1991) | {0, 1} |
| | `pol.cat.fl1` | Autocracy (Fearon and Laitin) | {0, 1} |
| | `pol.cat.fl2` | Anocracy (Fearon and Laitin) | {0, 1} |
| | `pol.cat.fl3` | Democracy (Fearon and Laitin) | {0, 1} |
| | `pol.cat.fl7` | Other (Fearon and Laitin) | {0, 1} |
| | `pol.durable.ln` | Regime duration, logged (Polity) | [0, 5.332719] |
| | `dis.l4pop.ln` | Percent of population subjected to state-led discrimination, logged | [0, 4.49981] |
| | `elf.ethnicc1` | Ethnic fractionalization: low | {0, 1} |
| | `elf.ethnicc2` | Ethnic fractionalization: medium | {0, 1} |
| | `elf.ethnicc3` | Ethnic fractionalization: high | {0, 1} |
| | `elf.ethnicc9` | Ethnic fractionalization: missing | {0, 1} |
| | `elc.eleth1` | Salient elite ethnicity: majority rule | {0, 1} |
| | `elc.eleth2` | Salient elite ethnicity: minority rule | {0, 1} |
| | `elc.eliti` | Ruling elites espouse an exclusionary ideology | {0, 1} |
| | `cou.tries5d` | Any coup attempts in past 5 years ($(t-4)$ to $(t)$) | {0, 1} |
| | `pit.sftpuhvl2.10.ln` | Sum of max annual magnitudes of PITF instability other than genocide from past 10 yrs ($(t-9)$ to $(t)$), logged | [0, 4.51086] |
| | `mev.regac.ln` | Scalar measure of armed conflict in geographic region, logged | [0, 4.174387] |
| | `mev.civtot.ln` | Scale of violent civil conflict, logged | [0, 2.397895] |

**Table. S1. Description of variables used in real data application 1.** The data and description are from the Early Warning Project (*2*)

| | Type I error avg (se) | Type I error violation rate | Type II error avg (se) |
|---|---|---|---|
| penLR (`default`) | 9.35% (0.14%) | 42% | 3.75% (0.07%) |
| penLR (`naïve`) | 20.53% (0.22%) | 94% | 1.25% (0.04%) |
| penLR (`NP`) | 4.48% (0.11%) | 7% | 6.47% (0.09%) |
| | | | |
| RF (`default`) | 12.45% (0.16%) | 66% | 5.72% (0.08%) |
| RF (`naïve`) | 53.15% (0.24%) | 100% | 0.34% (0.02%) |
| RF (`NP`) | 5.09% (0.12%) | 11% | 12.88% (0.12%) |
| | | | |
| NB (`default`) | 10.61% (0.14%) | 54% | 14.43% (0.11%) |
| NB (`naïve`) | 10.90 (0.17%) | 57% | 18.32% (0.62%) |
| NB (`NP`) | 1.87 (0.11%) | 4% | 76.27% (1.16%) |

**Table. S2. The performance of the NP umbrella algorithm in real data application 2.** Given $\alpha = 0.1$ and $\delta = 0.1$, after randomly splitting the data into training data with a size 374 (3/4 of the observations) and test data with a size 124 (1/4 of the observations) for 1000 times, we calculate the empirical type I and type II errors on the test data. The violation rates of empirical type I errors (the percentage of empirical type I errors greater than $\alpha$) of different approaches are summarized in the 2nd column. The averages and standard errors of empirical type II errors in percentages are summarized in the 3rd column. Three classification methods: penalized logistic regression (penLR), random forests (RF), and naïve Bayes (NB) are considered here. For each method, we considered three ways to choose the cutoff: `default`, the classical approach to minimize the overall classification error; `naïve`, the naïve approach to choose the empirical type I error on the training data to $\alpha$; `NP`, the NP umbrella approach.